\documentclass[twocolumn,conference,letterpaper]{IEEEtran}
\usepackage[left=0.75in,right=0.75in,top=1in,bottom=0.75in]{geometry}
\usepackage{comment}

\usepackage{cite}
\usepackage{amssymb,amsmath,latexsym,amsfonts,amsthm,mathtools}

\usepackage{graphicx}
\usepackage{float,subcaption}
\usepackage{textcomp}
\usepackage{xcolor}
\definecolor{darkpastelpurple}{rgb}{0.59, 0.44, 0.84}
\usepackage{hyperref}
\hypersetup{colorlinks, breaklinks, citecolor=blue, linkcolor=blue, urlcolor=blue}
\usepackage[nameinlink]{cleveref}
\Crefformat{figure}{#2Fig.~#1#3}
\Crefmultiformat{figure}{Figs.~#2#1#3}{ and~#2#1#3}{, #2#1#3}{ and~#2#1#3}

\crefname{assumption}{assumption}{assumptions}  
\Crefname{assumption}{Assumption}{Assumptions} 
\usepackage{tikz}
\usetikzlibrary{automata, shapes, arrows, calc, arrows.meta, fit, positioning}

\theoremstyle{plain}
\newtheorem{theorem}{Theorem}
\newtheorem{lemma}{Lemma}

\newtheorem*{problem*}{Problem}
\theoremstyle{remark}
\newtheorem{remark}{Remark}
\newtheorem{assumption}{Assumption}

\theoremstyle{definition}

\usepackage{mathrsfs}

\usepackage{newtxmath}
\usepackage{booktabs}
\usepackage{duckuments}

\IEEEoverridecommandlockouts

\begin{document}
	\title{Cooperative Integrated Estimation–Guidance for Simultaneous Interception of Moving Targets}
	\author{Lohitvel Gopikannan, Shashi Ranjan Kumar,~\IEEEmembership{Senior Member, IEEE}, and Abhinav Sinha,~\IEEEmembership{Senior Member,~IEEE}
		\thanks{L. Gopikannan and S. R. Kumar are with the Intelligent Systems \& Control (ISaC) Lab, Department of Aerospace Engineering, Indian Institute of Technology Bombay, Mumbai 400076, India. (e-mails: 24m0023@iitb.ac.in, srk@aero.iitb.ac.in). A. Sinha is with the Guidance, Autonomy, Learning, and Control for Intelligent Systems (GALACxIS) Lab, Department of Aerospace Engineering and Engineering Mechanics, University of Cincinnati, OH 45221, USA. (e-mail: abhinav.sinha@uc.edu).}
	}

	\maketitle
	\thispagestyle{empty}
	
	\begin{abstract}

This paper proposes a cooperative integrated estimation-guidance framework for simultaneous interception of a non-maneuvering target using a team of unmanned autonomous vehicles, assuming only a subset of vehicles are equipped with dedicated sensors to measure the target's states. Unlike earlier approaches that focus solely on either estimation or guidance design, the proposed framework unifies both within a cooperative architecture. To circumvent the limitation posed by heterogeneity in target observability, sensorless vehicles estimate the target's state by leveraging information exchanged with neighboring agents over a directed communication topology through a prescribed-time observer. The proposed approach employs true proportional navigation guidance (TPNG), which uses an exact time-to-go formulation and is applicable across a wide spectrum of target motions. Furthermore, prescribed-time observer and controller are employed to achieve convergence to true target's state and consensus in time-to-go within set predefined times, respectively. Simulations demonstrate the effectiveness of the proposed framework under various engagement scenarios.

	\end{abstract}
	
	
	\section{Introduction}\label{sec:introduction}
Simultaneous interception missions using teams of unmanned autonomous vehicles across aerial, terrestrial, and maritime domains have become a major focus in modern defense research. Traditional independently guided vehicles lack coordination, often resulting in dispersed arrivals and reduced mission effectiveness. In contrast, cooperative salvo guidance enables networked and adaptive decision-making, enhancing interception accuracy and resilience while offering improved modularity, scalability, and cost efficiency.

Early research in cooperative guidance focused on achieving the simultaneous interception of stationary targets. For example, the work in \cite{sp11} introduced a distributed cooperative guidance law based on biased proportional navigation (PN), whereas that in \cite{sp12} reformulated the impact-time control objective into an equivalent range-tracking problem. To alleviate communication burdens, authors in \cite{sp7} proposed finite-time distributed protocols that relied solely on locally computed time-to-go estimates. Along similar lines, the study in \cite{sp15} presented a receding-horizon cooperative strategy in which each interceptor exchanged information only with its immediate neighbors to solve local, finite-horizon optimization problems. In parallel, the approaches in \cite{sp16,sp17} eliminated the need for an explicit time-to-go estimate through a two-stage framework. Although these methods successfully attained consensus under undirected communication topologies, their asymptotic convergence makes them less suitable for time-critical interception missions. In \cite{sp1}, a leader–follower cooperative guidance architecture was proposed, wherein PN-based time-to-go estimation was integrated with a super-twisting sliding mode control to ensure finite-time convergence for simultaneous interception of stationary targets, even under large initial heading errors. However, a fundamental limitation of such leader–follower schemes lies in their susceptibility to leader failure, which can critically compromise the overall mission reliability. Utilizing the same time-to-go estimate, authors in \cite{hp13} proposed a guidance strategy that incorporates acceleration constraints and is applicable to both directed and undirected network topologies.

Both studies, \cite{sp1} and \cite{hp6}, extended their guidance strategies to moving targets. However, interception performance can degrade because this method relies on predicting the target’s future position, which may introduce errors and reduce effectiveness. Although studies in \cite{hp11, hp12} developed salvo guidance strategies over undirected and directed graphs, their reliance on radial acceleration to achieve time-to-go synchronization in finite time hinders practical implementation. To tackle unknown target maneuvers, the work in \cite{hp10} proposed a guidance strategy leveraging asymptotic consensus over undirected graphs, whereas those in \cite{hp4,hp7} employed deviated pursuit guidance with an exact time-to-go estimate to achieve time-constrained interception of moving targets by regulating the deviation angle. The authors in \cite{hp8} employed a weighted consensus approach for time-to-go synchronization over a pseudo-undirected graph, enabling a team of interceptors to achieve simultaneous interception using deviated pursuit guidance. By carefully selecting edge weights, including allowable negative values, the method allows precise, simultaneous interception outside the limits of the convex hull of initial time-to-go estimates. To overcome the limitations of deviated pursuit against stationary targets, authors in \cite{hp5} employed TPNG for time-constrained interception, thereby extending to a broader class of targets. Subsequently, a cooperative scheme was devised to enable simultaneous interception, with the interception time either preassigned or determined collaboratively. The work in \cite{hp9} proposed a predefined-time, leaderless deviated pursuit guidance and TPNG strategies for simultaneous interception of non-accelerating targets. In time-critical or short-duration engagements, prescribed-time consensus is especially valuable, as it guarantees convergence within a user-specified interval.

To the best of the author's knowledge, most of the aforementioned works assume all the vehicles have complete target information. However, from a practical standpoint, equipping every vehicle with sensing modules (sensors) to directly measure the target's states is generally impractical due to factors such as cost, payload limitations, power consumption, and integration complexity. A more viable alternative is to designate only a subset of vehicles with sensing modules, while the others infer the target's information through cooperative estimation based on distributed data fusion and localized coordination mechanisms. The proposed cooperative architecture thus facilitates fully distributed autonomy for coordinated target interception while significantly reducing sensing redundancy and system overhead.

This paper addresses a cooperative simultaneous interception problem involving a group of autonomous vehicles (such as multirotors or unmanned surface vessels) guided by TPNG against a non-maneuvering target. A limited number of vehicles are equipped with target-sensing modules, while the others estimate the target's states using a prescribed-time observer utilizing information exchanged over a directed communication topology. Subsequently, a prescribed-time control scheme is developed to achieve consensus in time-to-go for simultaneous interception.

\section{Background and Problem Formulation}
    \begin{figure}[h!]
    \centering
        \includegraphics[width=0.8\linewidth]{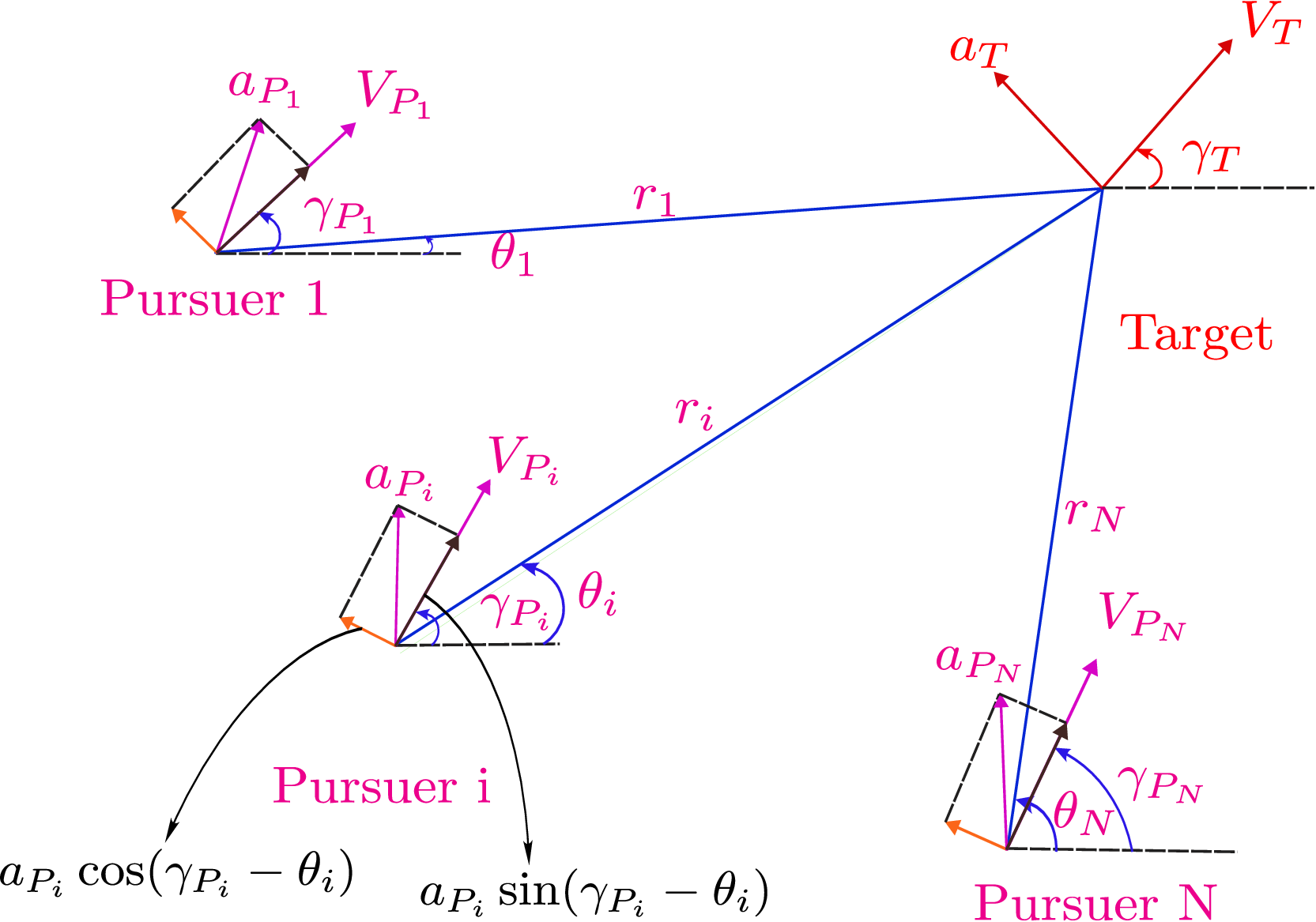}
        \caption{Multiple pursuers engaging a single target}
        \label{fig:enggeotpn}
    \end{figure}
In a cooperative multi-pursuer engagement scenario, a group of $N$ unmanned autonomous vehicles (pursuers) collaboratively engage a single target, as illustrated in \Cref{fig:enggeotpn}. The target moves with a constant velocity $V_{T}$, while the $i^{\text{th}}$ pursuer travels with velocity $V_{P_i}$. The engagement geometry is defined by the relative range $r_{i}$ and the line-of-sight (LOS) angle $\theta_{i}$ between the $i^{\text{th}}$ pursuer and the target. The flight path angle of the target is denoted by $\gamma_{T}$, whereas that of the $i^{\text{th}}$ pursuer is represented by $\gamma_{P_i}$. To maneuver and adjust its trajectory during the engagement, the $i^{\text{th}}$ pursuer generates an acceleration command $a_{P_i}$. The target and the pursuers are treated as ideal point-mass vehicles. During the design of the guidance strategy, the pursuers' dynamics may be ignored, assuming a sufficiently fast autopilot. The kinematic engagement equations are given by
\begin{subequations}
\begin{align}
\dot{r}_{i} &= V_{T}\cos\!\left(\gamma_{T} - \theta_{i}\right) 
              - V_{P_{i}}\cos\!\left(\gamma_{P_{i}} - \theta_{i}\right) 
              = V_{r_{i}},\\
r_{i}\,\dot{\theta}_{i} &= V_{T}\sin\!\left(\gamma_{T} - \theta_{i}\right) 
              - V_{P_{i}}\sin\!\left(\gamma_{P_{i}} - \theta_{i}\right) 
              = V_{\theta_{i}}, \\
\dot{\gamma}_{P_{i}} &= \frac{a_{P_{i}}\cos\!\left(\gamma_{P_{i}} - \theta_{i}\right)}{V_{P_{i}}}, \\
\dot{V}_{P_{i}} &= a_{P_{i}}\sin\!\left(\gamma_{P_{i}} - \theta_{i}\right).
\end{align}
\label{enggeo}
\end{subequations}
 \begin{assumption}\label{assum1}
The pursuer team consists of two classes of vehicles: one equipped with dedicated sensors capable of directly measuring the target's state, and the other relies on estimating that using information exchanged with neighboring vehicles over a directed graph $\mathcal{G}^s$.
 \end{assumption}
\begin{problem*}
    In accordance with \Cref{assum1}, design a cooperative integrated estimation-guidance strategy to ensure simultaneous interception of a non-maneuvering target.
\end{problem*}

As simultaneous interception requires precise control of each pursuer’s time-to-go, an accurate time-to-go expression is essential. Accordingly, we adopt TPNG as the baseline, which is applicable to wide spectrum of target motion and provides an accurate estimate of time-to-go \cite{hp5},
\begin{equation}\label{eq:tgo}
t_{\mathrm{go}_i} = - \frac{r_i \left(V_{r_i} + 2\,c_i \right)}{V_{\theta_i}^{2} + V_{r_i}^{2} + 2\,c_i\,V_{r_i}}, \quad 
c_i \gg \frac{V_{M_i} + V_{T_i}}{2}.
\end{equation}
\begin{lemma}
    For the $i^\text{th}$ pursuer, the dynamics of time-to-go \eqref{eq:tgo} has a relative degree of one with respect to its acceleration. 
\end{lemma}
\begin{proof}
On differentiating \eqref{eq:tgo} with respect to time and using \eqref{enggeo}, one may obtain
\begin{align}
\dot{t}_{\mathrm{go}_i} 
&= -1 
+ \frac{2 c_i (V_{r_i} + 2 c_i) (V_{\theta_i})^2}
       {\bigl((V_{\theta_i})^2 + (V_{r_i})^2 + 2 c_i V_{r_i}\bigr)^2} \notag \\
&- \frac{2 (V_{r_i} + 2 c_i) V_{\theta_i} r_i}
           {\bigl((V_{\theta_i})^2 + (V_{r_i})^2 + 2 c_i V_{r_i}\bigr)^2} a_{M_i},
\label{eq:tgodynamics}
\end{align}
indicating that $a_{P_i}$ appears in the first derivative of ${t}_{\mathrm{go}_i}$.
\end{proof}
\begin{remark}
    From \eqref{eq:tgodynamics}, it is evident that the acceleration command $a_{P_i}$ for regulating engagement duration requires knowledge of key engagement variables—specifically, the relative range and bearing to the target. These measurements, however, may not be available to all pursuers if some lack onboard sensors, thereby necessitating the requirement for the observer to estimate the target's states.
\end{remark}
The sensing topology is modeled by a directed graph $\mathcal{G^S} = \{\mathcal{V}, \mathcal{E}\}$, with node set $\mathcal{V} = \{v_0, v_1, \ldots, v_N\}$, where $v_0$ denotes the target and $v_1, \ldots, v_N$ represent the pursuers. The communication structure is captured by the directed edge set $\mathcal{E} \subseteq \mathcal{V} \times \mathcal{V}$, with $\varepsilon_{ij} \in \mathcal{E}$ denoting an information flow from $v_i$ to $v_j$. In-neighbor set for each agent $v_i \in \mathcal{V}$, is defined as $\mathcal{M}_i = \{v_j \in \mathcal{V} : \varepsilon_{ji} \in \mathcal{E}\}$, consisting of all agents that provide information to $v_i$. An agent is defined as a leader if it has no in-neighbors, and as a follower otherwise. In this framework, the target $v_0$ serves as the unique leader, broadcasting information without reception, and all interceptors act as followers. The overall communication structure is encoded in the adjacency matrix $\mathcal{W} = [w_{ij}] \in \mathbb{R}^{(N+1)\times(N+1)}$ where

\begin{equation*}
    w_{ij}=\begin{cases}
        1,&\text{if}~ e_{ji}\in \mathcal{E}\\
        0, & \text{otherwise}
    \end{cases}\quad\text{and}~a_{ii}=0.
\end{equation*}
The matrix $\mathcal{D} = \mathrm{diag}\{d_0, d_1, \ldots, d_N\}$ represents the in-degree matrix, where $d_i = \sum_{j=0}^{N} a_{ij}$ represents the total number of incoming edges associated with node $v_i$. The corresponding graph Laplacian is defined as $\mathcal{L} = \mathcal{D} - \mathcal{W}$, By means of the Kronecker product $(\otimes)$, scalar graph representations are generalized to describe higher-dimensional system dynamics.

\begin{assumption}[\cite{sp2}]\label{assum2}
Assuming that the sensing topology $\mathcal{G^S}$ contains a directed spanning tree with target ($v_0$) as the root node, then the corresponding Laplacian matrix is 
\begin{equation}
    \mathcal{L} = 
    \begin{bmatrix}
        0 & \mathbf{0}_{1\times N} \\
        \mathcal{L}_{EP} & \mathcal{L}_{PP}
    \end{bmatrix},
\end{equation}
where $\mathcal{L}_{EP} \in \mathbb{R}^{N \times 1}$ characterizes the influence exerted by the leader node (target) on each follower (pursuers) whereas $\mathcal{L}_{PP} \in \mathbb{R}^{N \times N}$ is the reduced Laplacian associated with the pursuer subgraph. 
\end{assumption}
\begin{lemma}
The eigenvalues of $ \mathcal{L}_{PP} $ satisfy $0 < \Re\{\lambda_1\} \leq \Re\{\lambda_2\} \leq \cdots \leq \Re\{\lambda_N\}$.     
\end{lemma}
\begin{lemma}[\cite{sp3}]\label{lem:ineq}
The matrix $\mathbf{\mathcal{L}}_{PP}$ is invertible (\Cref{assum2}). 
Consider $\mathbf{R} := \mathrm{diag}(\boldsymbol{\gamma})$, where 
$\boldsymbol{\gamma} = \left(\mathbf{\mathcal{L}}_{PP}^{\top}\right)^{-1} \mathbf{1}_N = [\gamma_1, \ldots, \gamma_N]^{\top}$.
Then, $\mathbf{R} \succ 0$ and $
\mathbf{Q}\left(\mathbf{\mathcal{L}}_{PP}\right) = 
\left( \mathbf{R}\mathbf{\mathcal{L}}_{PP} + \mathbf{\mathcal{L}}_{PP}^{\top}\mathbf{R} \right)\succ 0.$
\end{lemma}
Let $\mathcal{G^A} = (\mathcal{V^A}, \mathcal{E^A})$ denote the actuation graph, which specifies the interaction topology employed for coordinating guidance commands among the pursuers. The vertex set is $\mathcal{V}^A = {v_1^A, \dots, v_N^A}$, containing only pursuer nodes, while the edge set $\mathcal{E^A} \subseteq \mathcal{V^A} \times \mathcal{V^A}$ represents directed communication links that may differ from those in the sensing graph $\mathcal{G}^S$. In contrast to the sensing graph, $\mathcal{G^A}$ is assumed to be a leaderless, strongly connected digraph. The corresponding Laplacian matrix of $\mathcal{G}^A$ is denoted by $\mathcal{L}_P$.
\begin{remark}
    Employing distinct sensing and actuation graphs provides additional design flexibility, which enables independent tuning of the estimation and guidance modules while preserving the overall system stability and cooperative performance.
\end{remark}
\begin{lemma}[\cite{sp4}]\label{lem:mirror}
Consider a network of agents with a fixed interaction topology represented by a strongly connected digraph.  
Let ${\mathcal{{\hat L}}}$ denote the Laplacian matrix of the underlying undirected graph $\hat{\mathcal{G}}$ (i.e., the mirror of $\mathcal{G}$).  
Then, the network states $\mathbf{x}(t)$ satisfy $
\mathbf{x}^\top \mathcal{{L}} \mathbf{x} \geq \lambda_2(\mathcal{{\hat L}}) \|\mathbf{x}\|^2,
$
where $\lambda_2(\mathcal{{\hat L}})$ is the Fiedler eigenvalue of the mirror graph induced by $\mathcal{G}$.
\end{lemma}

Before proceeding further, we introduce a time-varying scaling function as follows:
\begin{align}\label{timevarying}
\varphi(t,T_c) =
\begin{cases}
\displaystyle \frac{T_c}{\pi} \sin\left(\frac{\pi}{T_c} t \right) + t - T_c, & 0 \le t < T_c, \\
1, & t \ge T_c,
\end{cases}
\end{align} and its derivative is obtained as \begin{align}
\dot{\varphi}(t,T_c) =
\begin{cases}
\displaystyle \cos\left(\frac{\pi}{T_c} t \right) + 1, & 0 \le t < T_c, \\
0, & t \ge T_c.
\end{cases}
\end{align}
where $t_c$ is a user-specified time. The trigonometric shaping functions \cite{sp5} utilized in \eqref{timevarying} ensure a smooth convergence. As \( \varphi(t, T_c) \) remains negative and \(\dot{\varphi}(t, T_c)\) stays positive yet decreases monotonically, approaching zero as \( t \to T_c^{-} \), the control laws are guaranteed to remain well-defined over the entire interval \([0, T_c)\).
\begin{lemma}[\cite{sp3}]\label{lem:ft}
Consider the dynamical system $\dot{\mathbf{x}}(t) = f(t, \mathbf{x}(t))$, where $\mathbf{x}(0) = \mathbf{x}_0(t) \in \mathbb{R}^n$ and $f: \mathbb{R}_+ \times \mathbb{R}^n \to \mathbb{R}^n$ is a nonlinear vector field that is locally bounded uniformly in time and let $V(x(t), t) : \mathcal{U} \times \mathbb{R}_+ \to \mathbb{R}$ 
be a continuously differentiable function and $\mathcal{U} \subset \mathbb{R}^n$ be a domain containing the origin. 
If there exists a real constant $b > 0$ such that
\begin{equation}
    V(0, t) = 0 \quad \text{and} \quad V(x(t), t) > 0 \ \text{in} \ \mathcal{U} \setminus \{0\},
\end{equation}
\begin{equation}
    \dot{V} \leq -bV - 2 \frac{\dot{a}}{a} V \quad \text{in} \ \mathcal{U}
\end{equation}
on $[t_0, \infty)$, then the origin of the system is prescribed-time stable with the prescribed time $T$ given by the user through the function $a(t)$ satisfying $\frac{\dot a (t)}{a(t)}>0$. If $\mathcal{U} = \mathbb{R}^n$, 
Then the origin of the system is globally prescribed-time stable with the prescribed time given by the function $a(t)$  . In addition, for 
$t \in [t_0, t_1)$, it holds that
\begin{equation}
    V(t) \leq a^{-2} \exp^{-b(t - t_0)} (a(t_0))^2V(t_0)
\end{equation}
and, for $t \in [t_1, \infty)$, it holds that $    V(t) \equiv 0.$
\end{lemma}
\section{Design of the Proposed Framework}
In this section, the design of a prescribed time distributed observer for estimating the target's state is presented first, followed by the development of a guidance strategy based on these estimated variables.

Let the target's state be defined as $\boldsymbol{\psi} = [x_T, y_T, \dot{x}_T, \dot{y}_T]^\top$, where $x_T$ and $y_T$ represent the target’s position, and $\dot{x}_T$ and $\dot{y}_T$ denote its velocity components in the inertial frame. Then the target's dynamics can be written as
\begin{align}
    \boldsymbol{\dot \psi}=\mathbf{A_T}\boldsymbol{\psi}=\begin{bmatrix}
        0 & 0 & 1&0\\
        0&0&0&1\\
        0&0&0&0\\
        0&0&0&0
    \end{bmatrix}\boldsymbol{\psi}.
\end{align}

The estimate of the target's state vector observed by the $i^\text{th}$ pursuer is denoted by $\boldsymbol{\hat{\psi}}$, with the corresponding relative estimation error defined as $\boldsymbol{\delta_i}$. Then the prescribed-time distributed observer to estimate $\boldsymbol{\psi}$ is given by
\begin{align}
\boldsymbol{\dot{\hat{\psi_i}}}(t) = \mathbf{A_T} \boldsymbol{\psi_i}
-\left(K_1 - K_2\frac{\dot \varphi(t,T_o)}{\varphi(t,T_o)}\right)\boldsymbol{\delta_i},
\label{eq:Stateobserver}
\end{align}
where $K_1$ and $K_2$ are  design parameters satisfying $K_1>\dfrac{2\|\mathbf{R}\otimes \mathbf{A_T}\|}{\lambda_{1}(\mathbf{Q})}$, $K_2\geq [\dfrac{2\lambda_{max}(\mathbf{R})}{\lambda_1(\mathbf{Q})}]$ and
\begin{align}
    \delta_i =& w_{i0} (\hat{\boldsymbol{\psi}}_i -{\boldsymbol{\psi}})+ \sum_{j=1}^N w_{ij} (\hat{\boldsymbol{\psi}}_i -\hat{\boldsymbol{\psi}}_j).
    \label{relative estimation error}
\end{align}
\begin{theorem}\label{thm:obs}
Consider the cooperative multi-pursuer–target engagement scenario described by \eqref{enggeo}, where only a subset of pursuers are equipped with sensors to directly measure the target's state. The $i^\text{th}$ pursuer employs the distributed observer \eqref{eq:Stateobserver} over a directed communication graph $\mathcal{G}^s$, under which, its state estimate, $\boldsymbol{\hat{\psi}}_i(t)$, converges to the true target's state $\boldsymbol{\psi}(t)$ within the prescribed time $T_o$, that is,
\begin{align}
\lim_{t \to T_o} \big( \boldsymbol{\hat{\psi}}_i(t) - \boldsymbol{\psi}(t) \big) &= 0.\label{observerstab}
\end{align}
regardless of the initial estimate $\boldsymbol{\hat{\psi}}_i(0)$.
\end{theorem}
\begin{proof}
Let \(\hat{\boldsymbol{\psi}} = [\hat{\boldsymbol{\psi}}_{1}^\top, \ldots, \hat{\boldsymbol{\psi}}_{N}^\top]^\top\) and \(\boldsymbol{\delta} = [\boldsymbol{\delta_{1}}^\top, \ldots, \boldsymbol{\delta_{N}}^\top]^\top\). For the $i^\text{th}$ pursuer, estimation error can be defined as  \(\tilde{{\boldsymbol{\psi}}}_{i} = \hat{\boldsymbol{\psi}}_{i} - {\boldsymbol{\psi}}\) then \(\tilde{{\boldsymbol{\psi}}} = [\tilde{\boldsymbol{\psi}}_{1}^\top, \ldots, \tilde{\boldsymbol{\psi}}^\top]^\top\). Then, using \eqref{eq:Stateobserver}, 
\begin{equation}
\dot{\tilde {\boldsymbol{\psi}}} = (\mathbf{I}_N \otimes \mathbf{A}_{T})\tilde {\boldsymbol{\psi}} - (\mathbf{I}_N \otimes \mathbf{I}_{4})\left(K_1-K_2\frac{\dot \varphi(t,T_o)}{\varphi(t,T_o)}\right)\boldsymbol{\delta}.
\end{equation}
From \eqref{relative estimation error}, one has \(\boldsymbol{\delta} = (\mathcal{L}_{PP} \otimes \mathbf{I}_{4}) \tilde{\boldsymbol{\psi}}\), whose dynamics is
\begin{equation}
        \dot{\boldsymbol{\delta}} =(\mathbf{I}_N \otimes \mathbf{A}_{T})\boldsymbol{\delta} - (\mathcal{L}_{PP} \otimes \mathbf{I}_{4})\left(K_1-K_2\frac{\dot \varphi(t,T_o)}{\varphi(t,T_o)}\right)\boldsymbol{\delta}.
\label{errorequation}
\end{equation}
Consider an auxiliary function $a_o(t)$ such that $a_o(t)=\dfrac{1}{\varphi(t,T_o)}$. On differentiating $a_o(t)$ with respect to time, we get
\begin{equation}
    \frac{\dot a_o(t)}{a_o(t)}=-\frac{\dot \varphi(t,T_o)}{\varphi(t,T_o)}.
\end{equation}
For all $t \in [0, T_o)$, the inequality $\dfrac{\dot \varphi(t,T_o)}{\varphi(t,T_o)} < 0$ holds, which implies that $\dfrac{\dot a_o(t)}{a_o(t)} > 0$ over the same interval. Therefore, \eqref{errorequation} becomes
\begin{equation}
        \dot{\boldsymbol{\delta}} =(\mathbf{I}_N \otimes \mathbf{A}_{T})\boldsymbol{\delta} - (\mathcal{L}_{PP} \otimes \mathbf{I}_{4})\left(K_1+K_2\frac{\dot a_o(t)}{a_o(t)}\right)\boldsymbol{\delta}.\label{errorequation1}
\end{equation}
Now, let us consider a Lyapunov function candidate 
\begin{align}
V_o = \boldsymbol{\delta}^{\top} (\mathbf{R} \otimes \mathbf{I}_4) \boldsymbol{\delta} \label{Lyapunov}
\end{align}
whose time derivative is
\begin{align}
\dot{V}_o
= 2 \boldsymbol{\delta}^{\top} (\mathbf{R} \otimes \mathbf{I}_4) \dot{\boldsymbol{\delta}}.
\end{align}
Using \eqref{errorequation1}, the above expression can be simplified to
\begin{align}
\dot{V}_o &= 2 \boldsymbol{\delta}^{\top} (\mathbf{R} \otimes \mathbf{I}_4) (\mathbf{I}_N \otimes \mathbf{A}_T) \boldsymbol{\delta} \notag \\
&\quad - 2 \boldsymbol{\delta}^{\top} (\mathbf{R} \otimes \mathbf{I}_4) (\mathcal{L}_{PP} \otimes \mathbf{I}_4) \left( K_1 + K_2 \frac{\dot a_o(t)}{a_o(t)} \right) \boldsymbol{\delta}\\
&= 2\boldsymbol{\delta}^{\top} \left(\mathbf{R}\otimes \mathbf{A}_T\right) \boldsymbol{\delta} \notag\\
&\quad - \left(K_1+K_2\frac{\dot a_o(t)}{a_o(t)}\right) 
\boldsymbol{\delta}^{\top} \left[ (\mathbf{R} \mathcal{L}_{PP} + \mathcal{L}_{PP}^{\top} \mathbf{R}) \otimes \mathbf{I}_4 \right] \boldsymbol{\delta}. \label{eq:vdotbeflem}
\end{align}
Upon using the results in \Cref{lem:ineq}, the expression in \eqref{eq:vdotbeflem} becomes
\begin{align}
\dot{V}_o&= 2\boldsymbol{\delta}^{\top} \left(\mathbf{R}\otimes \mathbf{A}_T\right) \boldsymbol{\delta} \nonumber\\
&- \left(K_1+K_2\frac{\dot a_o(t)}{a_o(t)}\right) \boldsymbol{\delta}^{\top} \left(\mathbf{Q}(\mathcal{L}_{PP}) \otimes \mathbf{I}_4 \right) \boldsymbol{\delta}, \notag
\end{align}
which can be further simplified based on the Rayleigh inequality and results from \Cref{lem:mirror} to
\begin{align}
\dot{V}_o\leq & 2 \left\|\mathbf{R}\otimes \mathbf{A}_T\right\| \boldsymbol{\delta}^{\top} \boldsymbol{\delta} - K_1 \lambda_1(\mathbf{Q}) \boldsymbol{\delta}^{\top}\boldsymbol{\delta}\nonumber \\&- K_2 \frac{\dot a_o(t)}{a_o(t)} \lambda_1(\mathbf{Q}) \boldsymbol{\delta}^{\top}\boldsymbol{\delta} \notag\\
\leq& - \frac{K_1 \lambda_1(\mathbf{Q}) - 2 \left\|\mathbf{R}\otimes \mathbf{A}_T\right\|}{\lambda_{\max}(\mathbf{R})} V_o - K_2 \frac{\dot a_o(t)}{a_o(t)} \frac{\lambda_1(\mathbf{Q})}{\lambda_{\max}(\mathbf{R})} V_o. \label{vdott}
\end{align}
For $K_1>\dfrac{2\|\mathbf{R}\otimes \mathbf{A_T}\|}{\lambda_{1}(\mathbf{Q})}$, $K_2\geq \dfrac{2\lambda_{\max}(\mathbf{R})}{\lambda_1(\mathbf{Q})}$ and let $\hat K_1=\dfrac{K_1\lambda_1(\mathbf{Q})-2\|\mathbf{R}\otimes \mathbf{A_T}\|}{\lambda_{\max}(\mathbf{R})}>0$,  \eqref{vdott} becomes,
\begin{align}
    \dot V_o\leq -\hat K_1 V_o-2\frac{\dot a_o(t)}{a_o(t)}V_o .
\end{align}
From \Cref{lem:ft}, it follows that
\begin{align}\label{vinequality2}
 a_o(t)^2 V_o(t) &\leq \exp\left( -\hat K_1t \right) a(0)^2 V_o(0) \notag \\
&= \frac{1}{T_o^2}\exp\left( -\hat{K_1}t \right)V_o(0) , 
\end{align}
which further implies that
\begin{align}
\|\boldsymbol{\delta}(t)\|^2 &\leq \frac{a_o(t)^{-2}}{T_o^2} \exp(-\hat K_1t) \frac{\lambda_{\max}(\mathbf{R})}{\lambda_{\min}(\mathbf{R})} \|\boldsymbol{\delta}(0)\|^2 , 
\end{align}
and then that
\begin{align}
\|\tilde{\boldsymbol{\psi}}(t)\| &= \| (\mathcal{L}_{PP} \otimes \mathbf{I}_{4})^{-1} \boldsymbol{\delta}(t) \| \leq \| (\mathcal{L}_{PP} \otimes \mathbf{I}_{4})^{-1} \| \cdot \|\boldsymbol{\delta}(t)\| \notag \\
&\leq \frac{1}{T_o |a_o(t)|} \exp(-\hat K_1t) 
      \sqrt{ \frac{\lambda_{\max}({\mathbf{R}})}{\lambda_{\min}(\mathbf{R})}} \notag\\
     &\times \| (\mathcal{L}_{PP} \otimes \mathbf{I}_{4})^{-1} \|\; \|\boldsymbol{\delta}(0)\| ,
\end{align}
which implies \eqref{observerstab}. This completes the proof.
\end{proof}

The $i^\text{th}$ pursuer, based on its estimated target's state $\hat{\psi}_i$, computes the estimated engagement variables, namely range $\hat{r}_i$ and LOS angle $\hat{\theta}_i$, as 
\begin{align}
    \hat r_i &=\sqrt{(\hat x_{T_i}-x_{P_i})^2+(\hat y_{T_i}-y_{P_i})^2},\\
    \hat \theta_i&=\tan^{-1}\left(\frac{\hat y_{T_i}-y_{P_i}}{\hat x_{T_i}-x_{P_i}}\right),
\end{align}
and the corresponding target's heading angle and velocity are estimated as
\begin{align}
   \hat \gamma_{T_i}&=\tan^{-1}{\dfrac{\dot y_{T_i}}{\dot x_{T_i}}},~\hat V_{T_i}=\sqrt{(\dot x_{T_i})^2+(\dot y_{T_i})^2}.
\end{align}
The modified engagement kinematics for an $i^\text{th}$
pursuer-target pair can now be given in terms of the estimated target's state $\boldsymbol{\hat \psi_i}$, which is
\begin{subequations}
\begin{align}
\hat V_{r_{i}} &= \hat V_{T_i}\cos\!\left(\hat \gamma_{T_i} - \hat \theta_{i}\right) 
              - V_{P_{i}}\cos\!\left(\gamma_{P_{i}} - \hat \theta_{i}\right), \\
 \hat V_{\theta_{i}}&= \hat V_{T_i}\sin\!\left(\hat \gamma_{T_i} - \hat\theta_{i}\right) 
              - V_{P_{i}}\sin\!\left(\gamma_{P_{i}} -\hat \theta_{i}\right) ,
            \\
\dot{\hat{\gamma}}_{P_i} &= \frac{a_{P_{i}}\cos\!\left(\gamma_{P_{i}} - \hat \theta_{i}\right)}{V_{P_{i}}}, \\
\dot{\hat{V}}_{P_i} &= a_{P_{i}}\sin\!\left(\gamma_{P_{i}} - \hat \theta_{i}\right).
\end{align}
\label{eq:enggeohat}
\end{subequations}
Define the error in achieving a common interception time for the $i^\text{th}$ pursuer as
\begin{equation}\label{timerror}
    e_i = t_{\mathrm{go}_i} - t_{\mathrm{go}}^d,
\end{equation}
where $t_{\mathrm{go}}^d$ is the desired time-to-go, determined cooperatively by the pursuers during the engagement rather than specified a priori. 
\begin{remark}
    For simultaneous interception, the distributed guidance law must drive the pursuers’ time-to-go values to consensus which is equivalent to achieving agreement in the error variables $e_i$, since for any pair $(i,j)$, $t_{\mathrm{go}_i} - t_{\mathrm{go}_j} = e_i - e_j$. Convergence of $e_i$ to zero guarantees that all pursuers intercept the target simultaneously.
\end{remark}
The following theorem presents the proposed guidance command for the $i^\text{th}$ pursuer that achieves this objective.
\begin{theorem}\label{amtheorem}
Consider the cooperative multi-pursuer–target engagement scenario described by \eqref{enggeo}, where only a subset of pursuers are equipped with sensors to directly measure the target's state and time-to-go \eqref{eq:tgo}. Over the actuation graph $\mathcal{G}^A$, each pursuer cooperatively exchanges its guidance command 
    \begin{align}
a_{P_i} &= 
c_i \dot{\hat{\theta}}_i
+ \frac{\left( (\hat{V}_{\theta_i})^2 + (\hat{V}_{r_i})^2 + 2 c_i \hat{V}_{r_i} \right)^2}
       {2 (\hat{V}_{r_i} + 2 c_i)\hat{V}_{\theta_i}\hat{r}_i} \notag \\
&\quad \times \left( M_1 - M_2 \frac{\dot \varphi(t,T_a)}{\varphi(t,T_a)} \right)
\sum_{j=1}^N \left[\mathcal{L}_P\right]_{ij}
\label{eq:am},
\end{align}
    for some $M_1>0,M_2 \geq \dfrac{1}{\lambda_2( {\mathcal{\hat L}_P})}$ to synchronize its time-to-go with its neighbors within a prescribed time $T_a > T_o << T_f$, thereby guaranteeing a simultaneous interception of the non-maneuvering target at a time $T_f$ cooperatively determined during the engagement.
\end{theorem}
\begin{proof}
The constant interception time, cooperatively agreed upon by all pursuers, satisfies $\dot{t}_{\mathrm{go}}^d = -1$. Differentiating \eqref{timerror} gives $\dot{e}_i = \dot{t}_{\mathrm{go}_i} + 1$. On substituting \eqref{eq:tgodynamics} into this expression, the resulting relation for the $i^{\text{th}}$ pursuer can be written as
\begin{align}
\dot{e}_i &=
\underbrace{ \frac{2 c_i (V_{r_i} + 2 c_i) V_{\theta_i}^2}
       {\left(V_{\theta_i}^2 + V_{r_i}^2 + 2 c_i V_{r_i}\right)^2} \ }_{F_i}
\,\underbrace{
- \frac{2 (V_{r_i} + 2 c_i) V_{\theta_i} r_i}
           {\left(V_{\theta_i}^2 + V_{r_i}^2 + 2 c_i V_{r_i}\right)^2}}_{B_i}a_{P_i}
\end{align}
Due to space constraints, the detailed steps are omitted.  Upon substituting the guidance command \eqref{eq:am} in the above equation  and simplifying, one can obtain
\begin{align*}
    \dot e_i&=F_i+B_i {a_P}_i= F_i+ \frac{B_i({u}_{c_i}-\hat F_i)}{\hat B_i},
\end{align*}
where
\begin{align*}
\hat{F}_i &= \frac{2 c_i (\hat{V}_{r_i} + 2 c_i) \hat{V}_{\theta_i}^2} {\left(\hat{V}_{\theta_i}^2 + \hat{V}_{r_i}^2 + 2 c_i \hat{V}_{r_i}\right)^2}, \quad
\hat{B}_i = -\frac{2 (\hat{V}_{r_i} + 2 c_i) \hat{V}_{\theta_i} r_i} {\left(\hat{V}_{\theta_i}^2 + \hat{V}_{r_i}^2 + 2 c_i \hat{V}_{r_i}\right)^2}, \\
{u}_{c_i} &= \left(-M_1+M_2\right)\frac{\dot{h}(t)}{h(t)}\sum_{j=1}^N[\mathcal{L}_P]_{ij} e_j.
\end{align*}
Let us define the error variables as $\tilde F_i= F_i-\hat F_i$ and $\tilde B_i= B_i-\hat B_i$ such that $\dot e_i$ becomes
\begin{align}
    \dot e_i &= F_i-\frac{B_i}{B_i- \tilde{B_i}}(F_i-\tilde F_i)+\frac{B_i}{B_i- \tilde{B_i}}{u_c}_i   
\end{align}
Define $\mu_i=\dfrac{\tilde B_i}{B_i}$ and $\omega_i=\dfrac{\tilde F_i}{F_i}$. Then, it follows that $\lim_{t\to T_o} \mu_i\to 0,~\lim_{t\to T_o}\omega_i\rightarrow 0$, and
\begin{align}
    \dot e_i=F_i\frac{\omega_i-\mu_i}{1-\mu_i}+\frac{{u_c}_i}{1-\mu_i}.
\end{align}
For notational simplicity, we express the variables in vector/matrix form. To this end, define
\begin{align}
    \mathbf{u}_c = &[u_{c_1}\;\; u_{c_2}\;\; \cdots\;\; u_{c_N}]^\top
= \left(-M_1+M_2\frac{\dot \varphi(t,T_a)}{\varphi(t,T_a)}\right)\mathcal{L}_P \mathbf{e}
\end{align}
with $\mathbf{e}=[e_1\;\; e_2\;\; \cdots\;\; e_N]^\top$. Similarly, $\boldsymbol{\omega} = [\omega_1 \; \cdots \; \omega_N]^\top$,  
$\boldsymbol{\mu} = [\mu_1 \; \cdots \; \mu_N]^\top$, and $\mathbf{F}=\mathrm{diag}(F_1\; \cdots \; F_N)$. Then, one also has
$    \bar{\mathbf{D}}_{\boldsymbol{\mu}} = 
\left[\cfrac{\omega_1-\mu_1}{1-\mu_1}\;\; \cdots\;\; \cfrac{\omega_N-\mu_N}{1-\mu_N}\right]^\top,~~\mathbf{D}_{\boldsymbol{\mu}} = 
\mathrm{diag}\!\left(\cfrac{1}{1-\mu_1}\;\; \cdots\;\; \cfrac{1}{1-\mu_N}\right)$ which leads us to write
\begin{align}
    \dot{\mathbf{e}} &= \mathbf{F}{\bar{\mathbf{D}}}_{\boldsymbol{\mu}} + \mathbf{D}_{\boldsymbol{\mu}}\mathbf{u}_e.\label{eq:ecl}
\end{align}
Considering a Lyapunov function candidate $V_c = \dfrac{1}{2}\mathbf{e}^\top\mathbf{e}$, we compute its time derivative as
\begin{align}
    \dot{V}_c &= \mathbf{e}^\top\dot{\mathbf{e}}
    = \mathbf{e}^\top\left(\mathbf{F}{\bar{\mathbf{D}}}_{\boldsymbol{\mu}} + \mathbf{D}_{\boldsymbol{\mu}}\mathbf{u}_e\right)\\ &= \mathbf{e}^\top\mathbf{F}{\bar{\mathbf{D}}}_{\boldsymbol{\mu}}
    + \mathbf{e}^\top\mathbf{D}_{\boldsymbol{\mu}}
    \left(-M_1 + M_2\frac{\dot{\varphi}(t,T_a)}{\varphi(t,T_a)}\right)\mathcal{L}_P\mathbf{e}.
\end{align}
We now analyze the stability of the closed-loop error dynamics \eqref{eq:ecl} to show that the system remains uniformly ultimately bounded before observer error convergence and achieves globally prescribed-time stability

Similar to the proof of \Cref{thm:obs}, we introduce an auxiliary function $a_c(t) = \dfrac{1}{\varphi(t,T_a)}$, which, on differentiating, gives 
$\dfrac{\dot{a_c}(t)}{a_c(t)} = -\dfrac{\dot{\varphi}(t,T_a)}{\varphi(t,T_a)} > 0$. Notice that 
$[\mathbf{D}_{\boldsymbol{\mu}}]_{ii}=\dfrac{1}{1-\mu_i}=\dfrac{B_i}{\hat B_i}
>0$. Define
$    d_{\min} \coloneqq \inf_{0 \le s < T_o} [\mathbf{D}_{\boldsymbol{\mu}}(s)]_{ii} > 0,\quad
d_{\max} \coloneqq \sup_{0 \le s < T_o} [\mathbf{D}_{\boldsymbol{\mu}}(s)]_{ii},
$ then $d_{\min}\mathbf{I}\le \mathbf{D}_{\boldsymbol{\mu}} \le d_{\max}\mathbf{I}$. 
Moreover, $\lim_{t\to T_o} \mathbf{D}_{\boldsymbol{\mu}}=\mathbf{I}\implies d_{\min}=d_{\max}=1$. Therefore,
\begin{align}
    \dot{V}_c &\le \mathbf{e}^\top\mathbf{F}{\bar{\mathbf{D}}}_{\boldsymbol{\mu}}
    - d_{\min}\mathbf{e}^\top\left(M_1+M_2\frac{\dot a_c(t)}{a_c(t)}\right)\mathcal{L}_P\mathbf{e}\\
    &= \mathbf{e}^\top\mathbf{F}{\bar{\mathbf{D}}}_{\boldsymbol{\mu}}
    - d_{\min}M_1\,\mathbf{e}^\top\mathbf{e}
    - d_{\min}M_2\frac{\dot a_c(t)}{a_c(t)}\,\mathbf{e}^\top\mathcal{L}_P\mathbf{e} \nonumber\\
    &\le \mathbf{e}^\top\mathbf{F}\mathbf{D}_{\boldsymbol{\mu}}
    \,{\mathbf{D}}_{\boldsymbol{\omega}}
    - 2d_{\min}M_1 V_c
    - 2d_{\min}M_2\frac{\dot a_c(t)}{a_c(t)}\lambda_2({\mathcal{\hat L}_P})V_c \nonumber\\
    &\le d_{\min}\left[\frac{d_{\max}}{d_{\min}}\,
    \mathbf{e}^\top\mathbf{F}\mathbf{D}_{\boldsymbol{\omega}}
    - 2M_1 V_c
    - 2M_2\frac{\dot a_c(t)}{a_c(t)}\lambda_2({\mathcal{\hat L}_P})V_c\right],\label{eq:V2dot}
\end{align}
where $\mathbf{D}_{\boldsymbol{\omega}}=[\omega_1 - \mu_1,\; \ldots,\; \omega_N - \mu_N]^\top$. For $M_2 \ge \dfrac{1}{\lambda_2({\mathcal{\hat L}_P})}$, letting $\hat{M_1}=2M_1$ allows us to simplify \eqref{eq:V2dot} to
\begin{align}
    \dot V_c &\le d_{\min}\left[\frac{d_{\max}}{d_{\min}}\,
    \mathbf{e}^\top\mathbf{F}\mathbf{D}_{\boldsymbol{\omega}}
    - \hat{M_1} V_c
    - 2\frac{\dot a_c(t)}{a_c(t)}V_c\right].\label{pre_v_dot}
\end{align}
Thereafter, note that $\|\mathbf{e}^\top\mathbf{F}\mathbf{D}_{\boldsymbol{\omega}}\|_2 
    \leq \|\mathbf{e}\|_2 \|\mathbf{F}\|_2 \|\mathbf{D}_{\boldsymbol{\omega}}\|_2  \leq \sqrt{\mathbf{e}^\top\mathbf{e}}\,\|\mathbf{F}\|_2 \|\mathbf{D}_{\boldsymbol{\omega}}\|_2 = \sqrt{2V_c}\,\|\mathbf{F}\|_2 \|\mathbf{D}_{\boldsymbol{\omega}}\|_2$. On using this result to simplify $\dot V_c$ further, one has
\begin{align}
    \dot{V}_c &\leq d_{\min}\left[
    \frac{d_{\max}}{d_{\min}}\sqrt{2V_c}\,\|\mathbf{F}\|_2\|\mathbf{D}_{\boldsymbol{\omega}}\|_2
    - \hat{M_1} V_c
    - 2\frac{\dot{a_c}(t)}{a_c(t)}V_c
    \right].\label{v1dot}
\end{align}
For $t \geq T_o$, it follows that $\|\mathbf{D}_{\boldsymbol{\omega}}\|_2 = 0$ and $d_{\min} = d_{\max} = 1$. Therefore, \eqref{v1dot} can be expressed as
\begin{align}
    \dot V_c&\leq -\hat M_1 V_c-2\frac{\dot a_e(t)}{a_e(t)}V_c,
\end{align}
and from \Cref{lem:ft}, it follows that
\begin{align}
    V_c& \leq \frac{\varphi^2(t,T_a)}{T_a^2}e^{-\hat{M_1} t} V_c(0) ,  \label{eq:vfinal}
\end{align}
and hence,
\begin{align}
    \|\mathbf{e}(t)\|_2 &\leq \frac{|\varphi(t,T_a)|}{T_a} e^{\frac{-\hat{M_1} t}{2}}\|\mathbf{e}(0)\|_2.\label{eq:efinal}
\end{align}
Note that due to space constraints, detailed intermediate steps to arrive at \eqref{eq:vfinal}--\eqref{eq:efinal} have been omitted. Based on \eqref{eq:vfinal}--\eqref{eq:efinal}, we conclude that $V_c(t) \equiv 0$ yields $\mathbf{e}(t) \equiv 0$ as $t \rightarrow T_a$. Combined with the results from \Cref{lem:ft}, this establishes that the closed-loop error dynamics is globally prescribed-time stable. Thus, a consensus in $e_i$ and hence time-to-go is achieved within $T_a$, which leads to a simultaneous interception of the target.
\end{proof}

\section{Performance Evaluation}

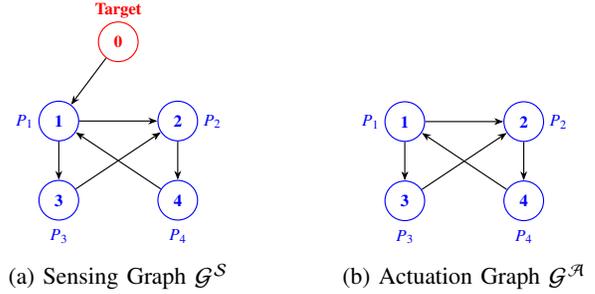
\begin{figure}[h!]
	\centering
	\begin{subfigure}[b]{0.47\linewidth}
		\centering
		\resizebox{.7\linewidth}{!}{%
			\begin{tikzpicture}[
				->, >=Stealth, thick,
				font=\bfseries\large  
				]
				
				\node[circle, draw=red, text=red, minimum size=1cm, inner sep=0pt] (0) at (2.5,4) {0};
				\node[circle, draw=blue, text=blue, minimum size=1cm, inner sep=0pt] (1) at (1,2) {1};
				\node[circle, draw=blue, text=blue, minimum size=1cm, inner sep=0pt] (2) at (4,2) {2};
				\node[circle, draw=blue, text=blue, minimum size=1cm, inner sep=0pt] (3) at (1,0) {3};
				\node[circle, draw=blue, text=blue, minimum size=1cm, inner sep=0pt] (4) at (4,0) {4};
				
				\draw[->] (0) -- (1);              
				\draw[->] (1) -- (2);              
				\draw[->] (1) -- (3);              
				\draw[->] (2) -- (4);              
				\draw[->] (3) -- (2);              
				\draw[->] (4) -- (1);              
				
				\node[text=red] at (2.5,4.8) {Target};
				\node[text=blue, anchor=east] at (1.west) {$P_1$};
				\node[text=blue, anchor=west] at (2.east) {$P_2$};
				\node[text=blue] at (1,-0.9) {$P_3$};
				\node[text=blue] at (4,-0.9) {$P_4$};
				
			\end{tikzpicture}%
		}
		\caption{Sensing Graph $\mathcal{G^S}$}
		\label{sensing_graphtpn}
	\end{subfigure}
	\hfill
	\begin{subfigure}[b]{0.47\linewidth}
		\centering
		\resizebox{.7\linewidth}{!}{%
			\begin{tikzpicture}[
				->, >=Stealth, thick,
				font=\bfseries\large  
				]
				\node[circle, draw=blue, text=blue, minimum size=1cm, inner sep=0pt] (1) at (1,2) {1};
				\node[circle, draw=blue, text=blue, minimum size=1cm, inner sep=0pt] (2) at (4,2) {2};
				\node[circle, draw=blue, text=blue, minimum size=1cm, inner sep=0pt] (3) at (1,0) {3};
				\node[circle, draw=blue, text=blue, minimum size=1cm, inner sep=0pt] (4) at (4,0) {4};
				\draw[->] (1) -- (2);              
				\draw[->] (1) -- (3);              
				
				\draw[->] (2) -- (4);              
				\draw[->] (4) -- (1);              
				\draw[->] (3) -- (2);              
				\node[text=blue, anchor=east] at (1.west) {$P_1$};
				\node[text=blue, anchor=west] at (2.east) {$P_2$};
				\node[text=blue] at (1,-0.9) {$P_3$};
				\node[text=blue] at (4,-0.9) {$P_4$};
				
			\end{tikzpicture}%
		}
		\caption{Actuation Graph $\mathcal{G^A}$}
		\label{actuation_graphtpn}
	\end{subfigure}
	\caption{Pursuers' communication topologies.}
	\label{fig:graphsTPN}
\end{figure}
The effectiveness of the proposed strategy is first validated through a simulation scenario involving a single constant-velocity target and four pursuers ($P_1$–$P_4$). The target moves with a speed of $50$ m/s and heading angle of $120^\circ$, starting from the position $(2500,0)$ m. Among the pursuers, only $P_1$ is equipped with a sensor to directly measure the target's states, while the remaining pursuers obtain target information through communication governed by the sensing topology $\mathcal{G^S}$ illustrated in \Cref{sensing_graphtpn}. The corresponding actuation topology $\mathcal{G^A}$ is shown in \Cref{actuation_graphtpn}, which is leaderless. All pursuers are launched from the origin with initial speeds of $\begin{bmatrix} 55 & 57 & 58 & 60 \end{bmatrix}$ m/s and heading angles of $\begin{bmatrix} 10^\circ & 15^\circ & 20^\circ & 25^\circ \end{bmatrix}$. Their initial estimates of the target’s position are
$\begin{bmatrix} (3050,500) & (4500,100) & (2500,200) & (3500,400) \end{bmatrix}$ m, while
$\begin{bmatrix} (25,25) & (20,40) & (40,15) & (25,25) \end{bmatrix}$ m/s is their  corresponding velocity estimates. From these estimates, the initial time-to-go values are obtained as 
$\begin{bmatrix}
107.359 & 135.118 & 169.242 & 112.380
\end{bmatrix}$ s, 
while the corresponding true time-to-go values are 
$\begin{bmatrix}
32.427 & 31.804 & 31.833 & 30.630
\end{bmatrix}$ s. Each pursuer is subject to actuator saturation, with the maximum allowable lateral acceleration limited to $7$ g, where g denotes gravitational acceleration. The observer and controller gains are selected according to the design criteria. The prescribed observer convergence time is set to $T_o = 0.6$ s, and the consensus controller convergence time is specified as $T_a = 3$ s. The value of $c_i$ is considered as $c_i=3(V_{P_i}+V_{T}).$
\begin{figure}
    \centering
    
    \begin{subfigure}{0.23\textwidth}
        \centering
        \includegraphics[width=\linewidth]{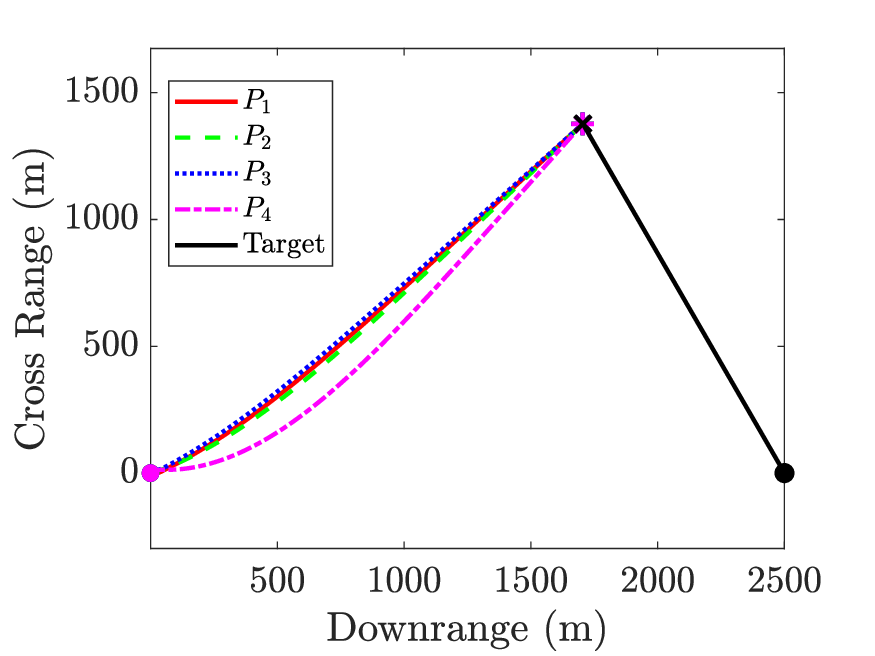}
        \caption{Trajectory.}
        \label{fig:Trajectorytpn1}
    \end{subfigure}
    \begin{subfigure}{0.23\textwidth}
        \centering
        \includegraphics[width=\linewidth]{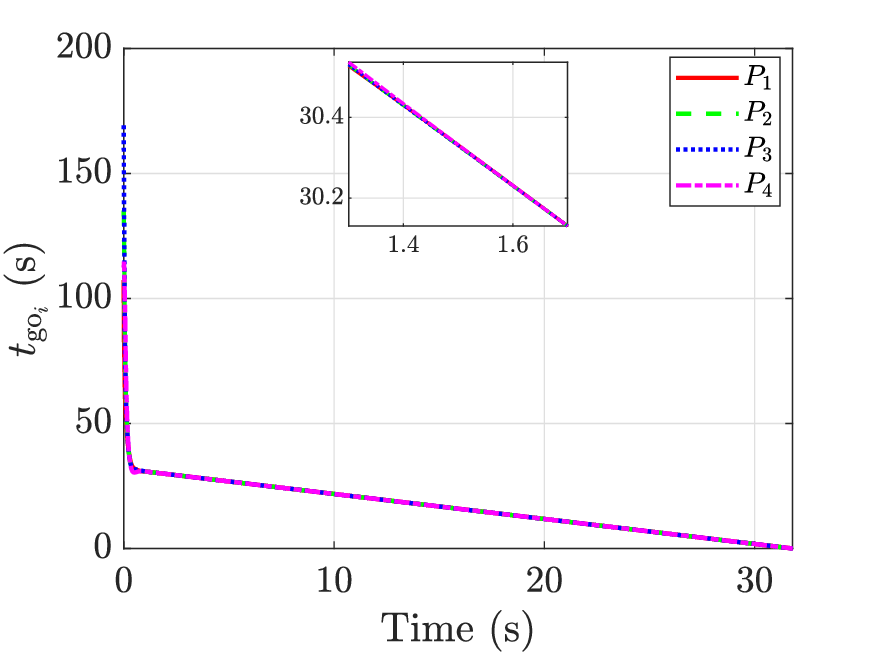}
        \caption{Time-to-go.}
        \label{fig:Time-to-gotpn1}
    \end{subfigure}
    
    \begin{subfigure}{0.23\textwidth}
        \centering
        \includegraphics[width=\linewidth]{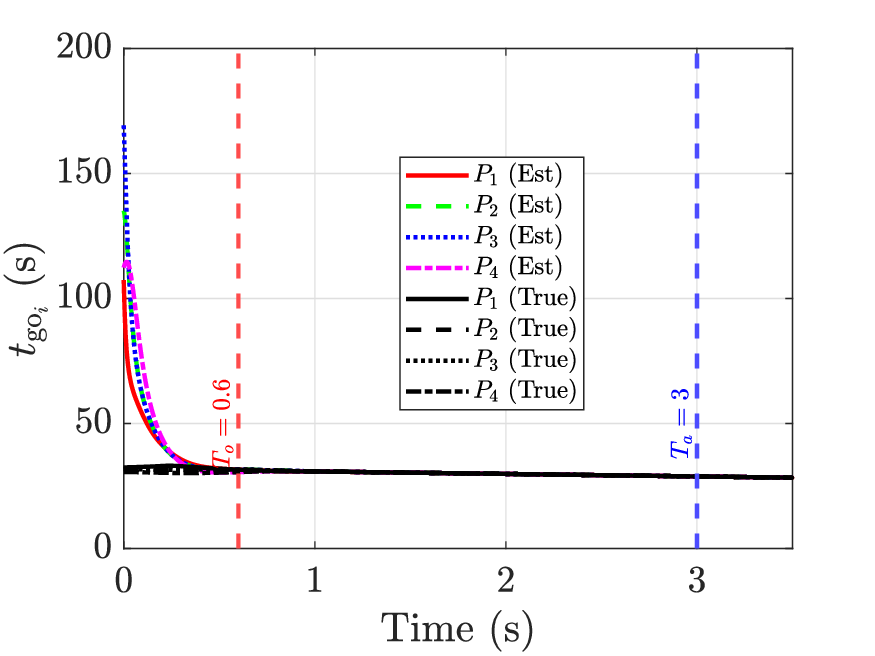}
        \caption{True vs estimated time-to-go.}
        \label{fig:fig:truevsestim_combinedtpn11}
    \end{subfigure}
    \begin{subfigure}{0.23\textwidth}
        \centering
        \includegraphics[width=\linewidth]{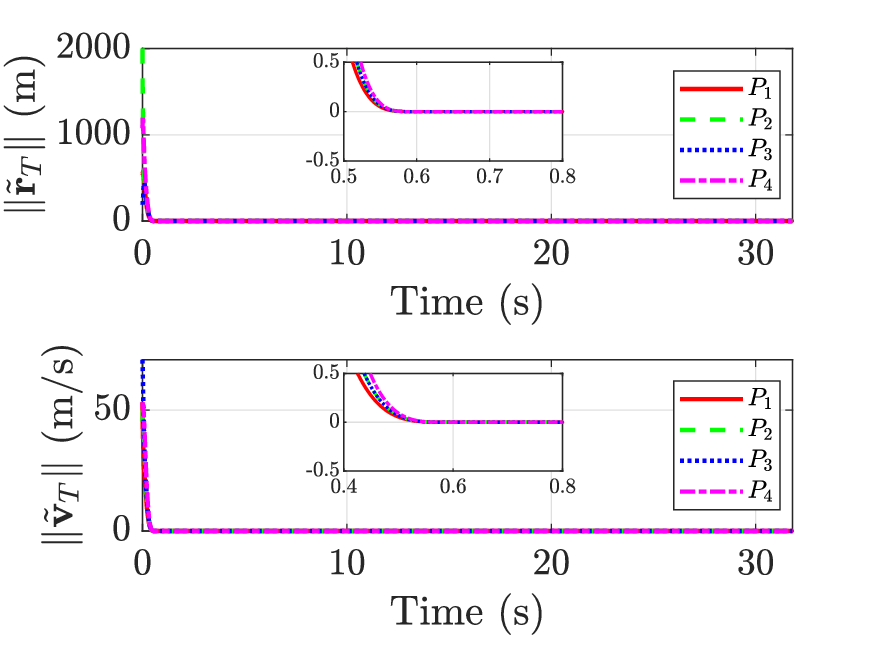}
        \caption{Observer estimation error.}
        \label{fig:Observer_errortpn1}
    \end{subfigure}
    
    \begin{subfigure}{0.23\textwidth}
        \centering
        \includegraphics[width=\linewidth]{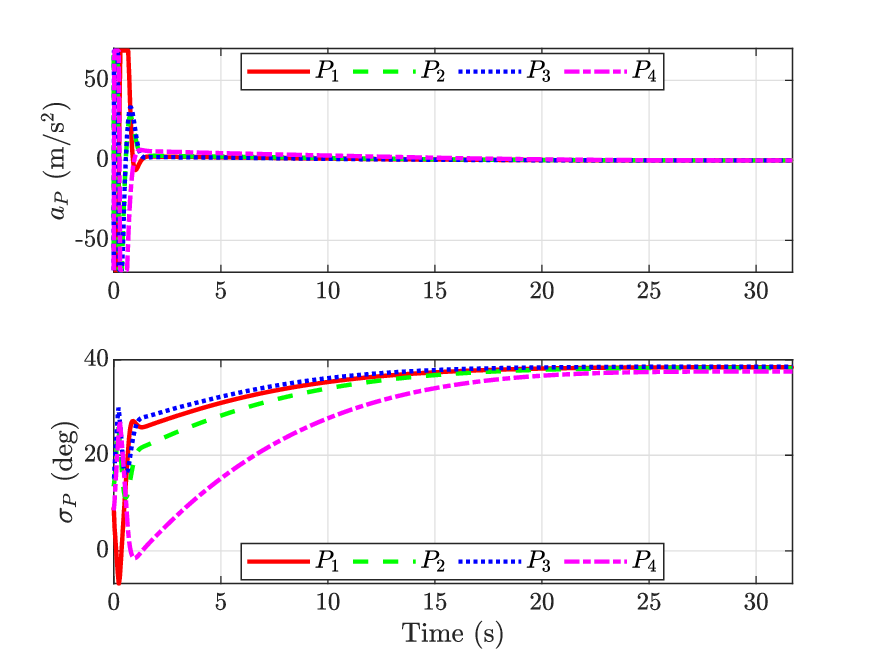}
        \caption{Acceleration \&l lead Angle.}
        \label{fig:accelerationtpn1}
    \end{subfigure}
    \begin{subfigure}{0.23\textwidth}
        \centering
        \includegraphics[width=\linewidth]{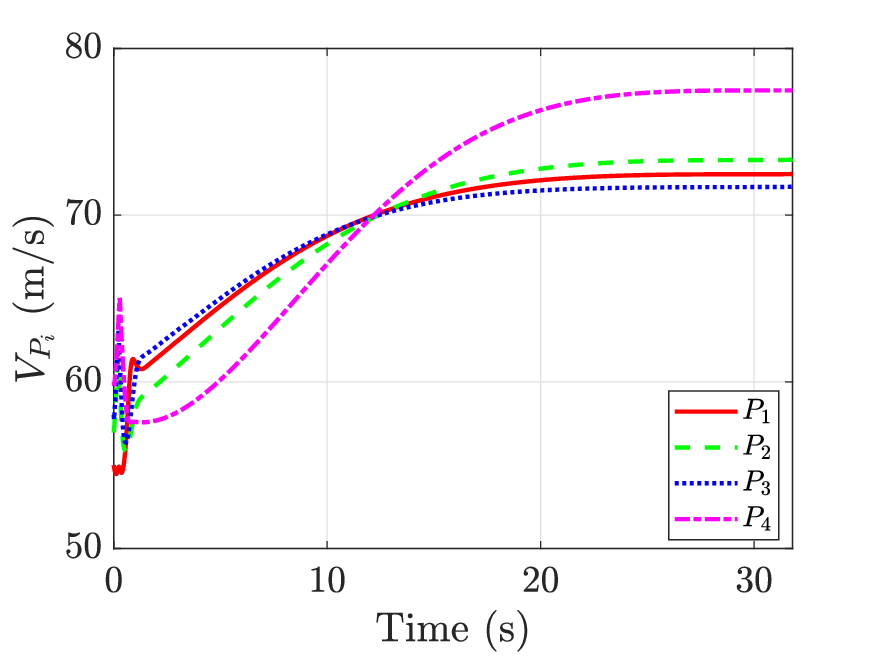}
        \caption{Velocity Evolution.}
        \label{fig:veltpn1}
    \end{subfigure}
    
    \caption{Performance evaluation of the proposed strategy for a constant velocity target.}
    \label{fig:tpn1}
\end{figure}

\Cref{fig:tpn1} illustrates the engagement scenario, in which pursuers accomplish cooperative simultaneous interception of a constant velocity moving target at $31.8$ s despite having incomplete information. As observed from \Cref{fig:Time-to-gotpn1}, this information asymmetry does not hinder coordination since the pursuers achieve consensus on a common time-to-go well within $1.5$ s, which is well below the prescribed convergence time of $T_a = 3$ s, thereby attaining the geometry required for simultaneous interception. The corresponding accelerations and lead angles are illustrated in \Cref{fig:accelerationtpn1}. It is evident that, before consensus on time-to-go, the lead angles exhibit fluctuations, and acceleration demands are relatively high. However, once the observer error converges to zero within the prescribed time $T_o = 0.6$ s (see \Cref{fig:Observer_errortpn1}), and consensus in time-to-go is attained, both the lead angles and accelerations settle to stable values. Consistent with the lead angle and acceleration responses, \Cref{fig:veltpn1} demonstrates that the velocity profile exhibits initial transients, which subside as the observer error converges to zero and consensus in time-to-go is established, resulting in a steady velocity phase coinciding with the constant lead angle regime.

\begin{figure}[h!]
    \centering  
    \begin{subfigure}{0.23\textwidth}
        \centering
        \includegraphics[width=\linewidth]{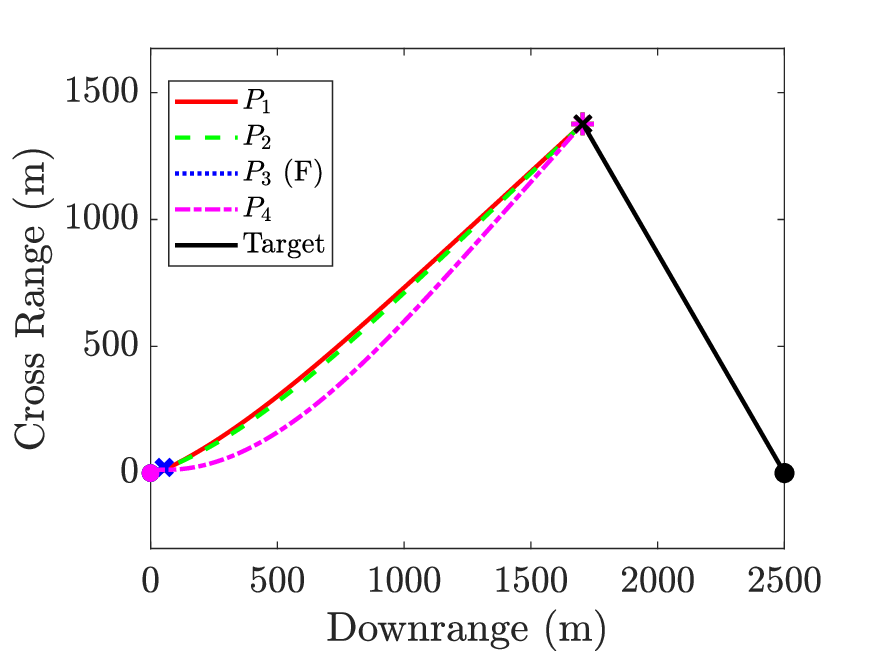}
        \caption{Trajectory.}
        \label{fig:Trajectorytpn3}
    \end{subfigure}
    \begin{subfigure}{0.23\textwidth}
        \centering
        \includegraphics[width=\linewidth]{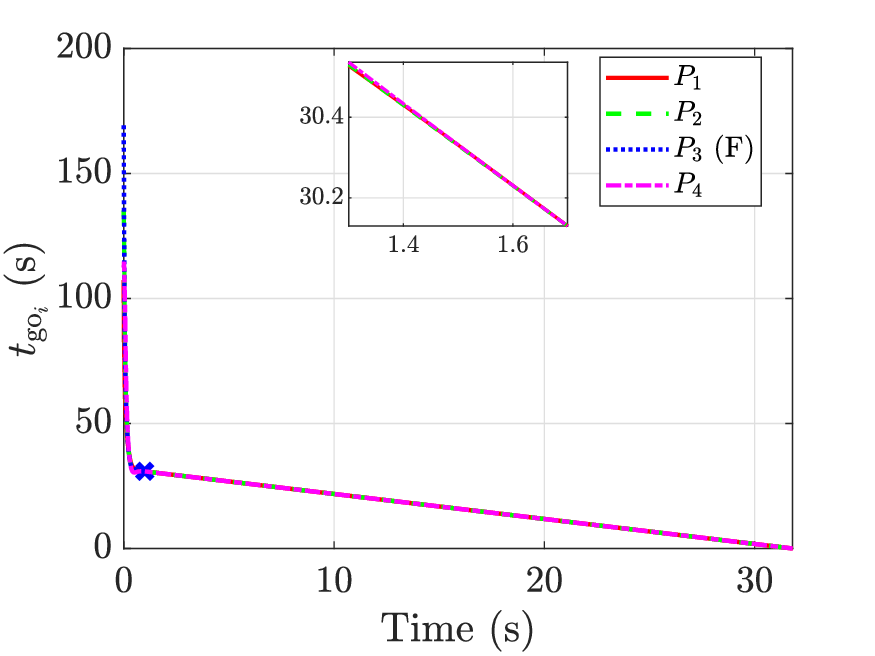}
        \caption{Time-to-go.}
        \label{fig:Time-to-gotpn3}
    \end{subfigure}
    
    \begin{subfigure}{0.23\textwidth}
        \centering
        \includegraphics[width=\linewidth]{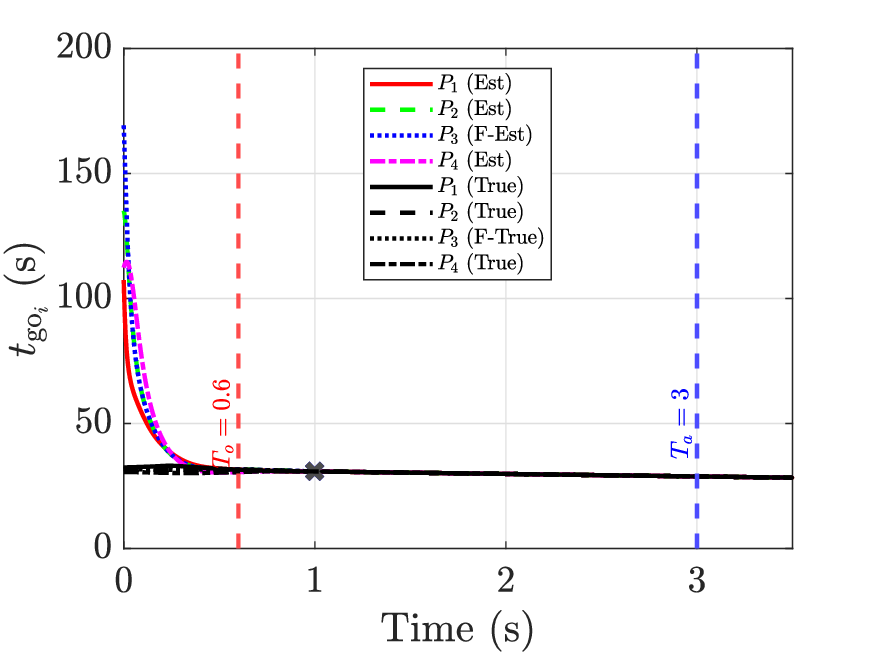}
        \caption{True vs Estimated Time-to-go.}
        \label{fig:fig:truevsestim_combinedtpn3}
    \end{subfigure}
    \begin{subfigure}{0.23\textwidth}
        \centering
        \includegraphics[width=\linewidth]{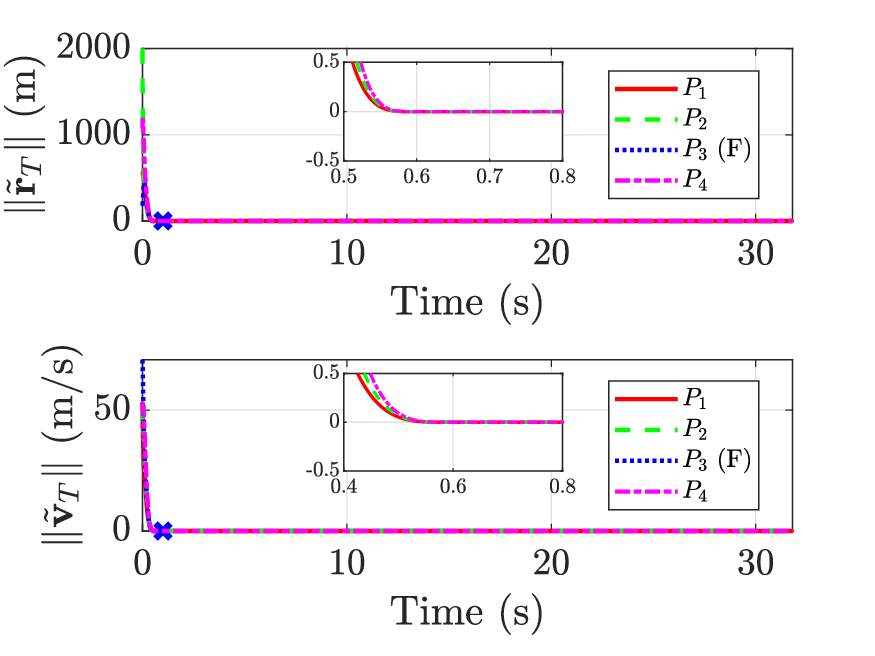}
        \caption{Observer Estimation Error.}
        \label{fig:Observer_errortpn3}
    \end{subfigure}
    
    \begin{subfigure}{0.23\textwidth}
        \centering
        \includegraphics[width=\linewidth]{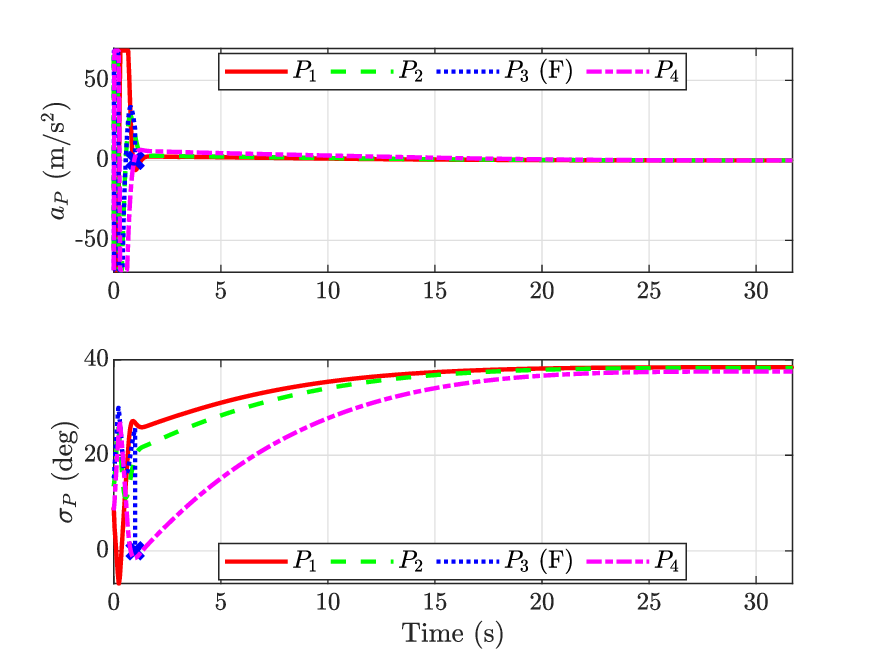}
        \caption{Acceleration \& Lead Angle.}
        \label{fig:accelerationtpn3}
    \end{subfigure}
    \begin{subfigure}{0.23\textwidth}
        \centering
        \includegraphics[width=\linewidth]{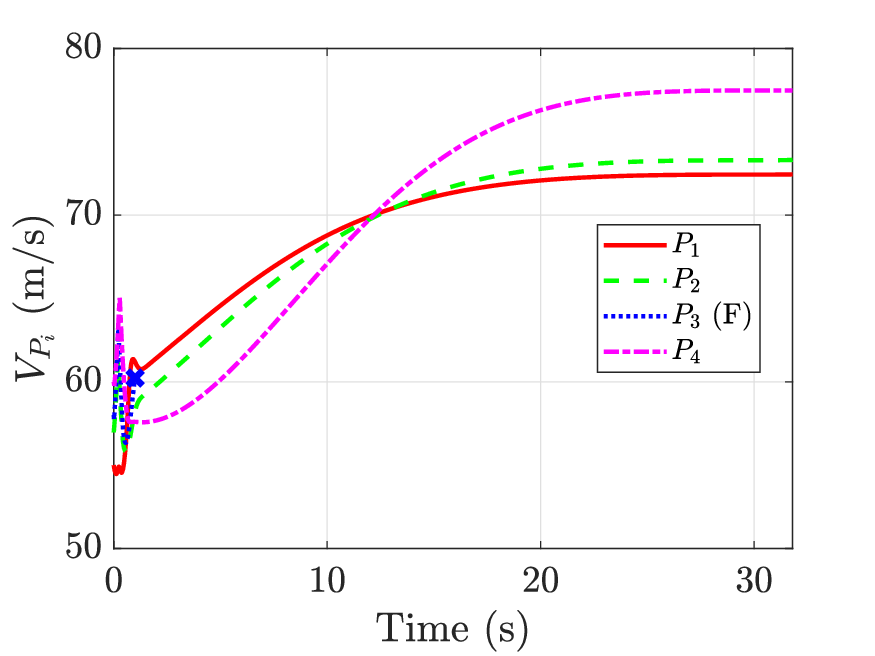}
        \caption{Velocity Evolution.}
        \label{fig:veltpn3}
    \end{subfigure}
    
    \caption{Performance evaluation of the proposed strategy when pursuer 3 fails.}
    \label{fig:tpn3}
\end{figure}

Consider the engagement scenario where pursuer~3 experiences a complete failure at \( t = 1 \) s while all other conditions remain unchanged. As illustrated in \Cref{fig:tpn3}, the remaining pursuers continue to operate cooperatively and accomplish simultaneous interception of the target at approximately \( 31.8 \) s. This outcome demonstrates the inherent robustness and resilience of the proposed cooperative framework despite the loss of one agent, where coordination and synchronization are preserved owing to the structural integrity of the communication and control topologies. Specifically, as long as the sensing graph \( \mathcal{G}^S \) retains a spanning tree and the actuation graph \( \mathcal{G}^A \) remains strongly connected, the cooperative guidance law ensures successful interception without any degradation in performance.

  \begin{figure}[h!]
    \centering  
    \begin{subfigure}{0.23\textwidth}
        \centering
        \includegraphics[width=\linewidth]{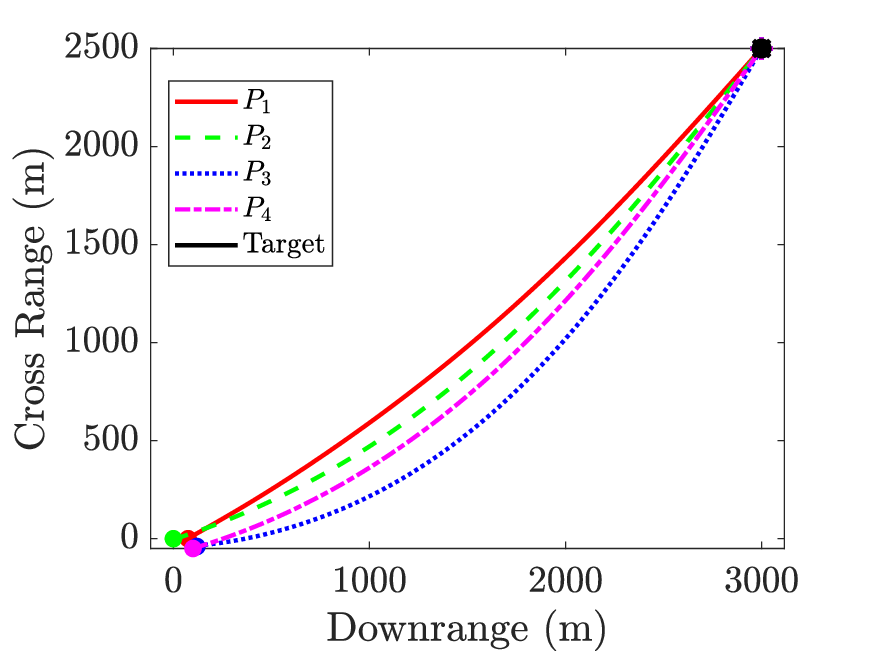}
        \caption{Trajectory.}
        \label{fig:Trajectorytpn2}
    \end{subfigure}
    \begin{subfigure}{0.23\textwidth}
        \centering
        \includegraphics[width=\linewidth]{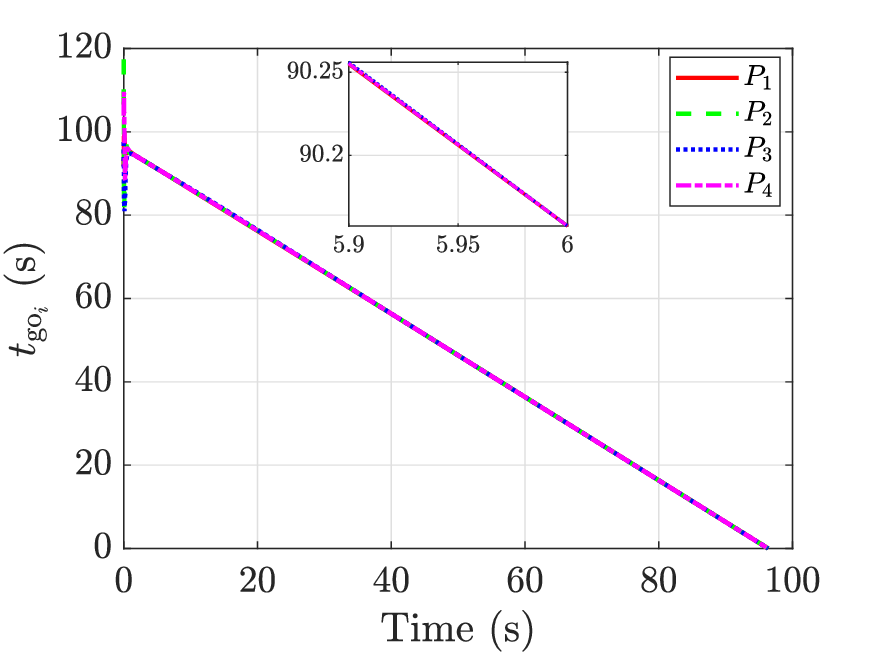}
        \caption{Time-to-go.}
        \label{fig:Time-to-gotpn2}
    \end{subfigure}
    
    \begin{subfigure}{0.23\textwidth}
        \centering
        \includegraphics[width=\linewidth]{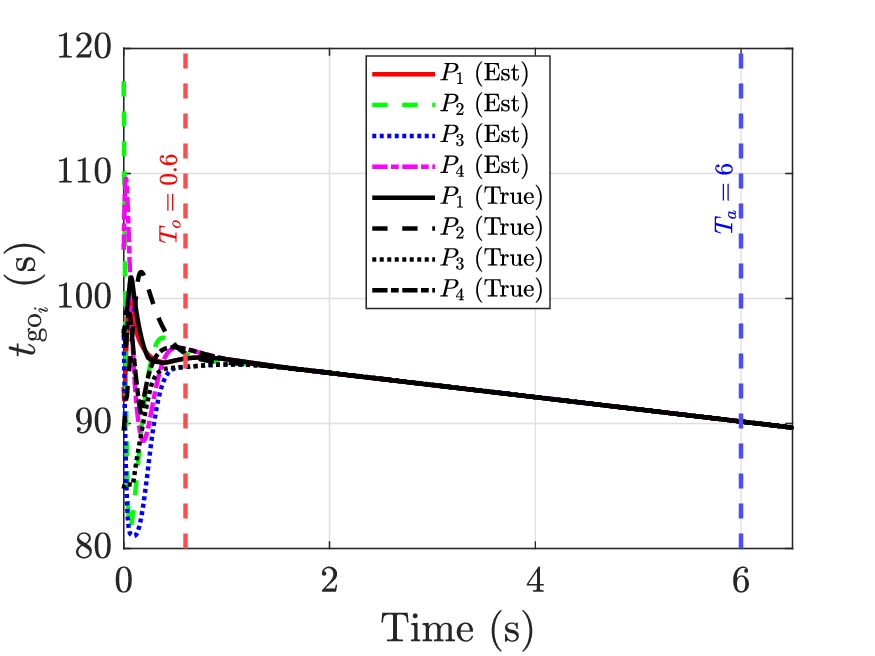}
        \caption{True vs Estimated Time-to-go.}
        \label{fig:fig:truevsestim_combinedtpn2}
    \end{subfigure}
    \begin{subfigure}{0.23\textwidth}
        \centering
        \includegraphics[width=\linewidth]{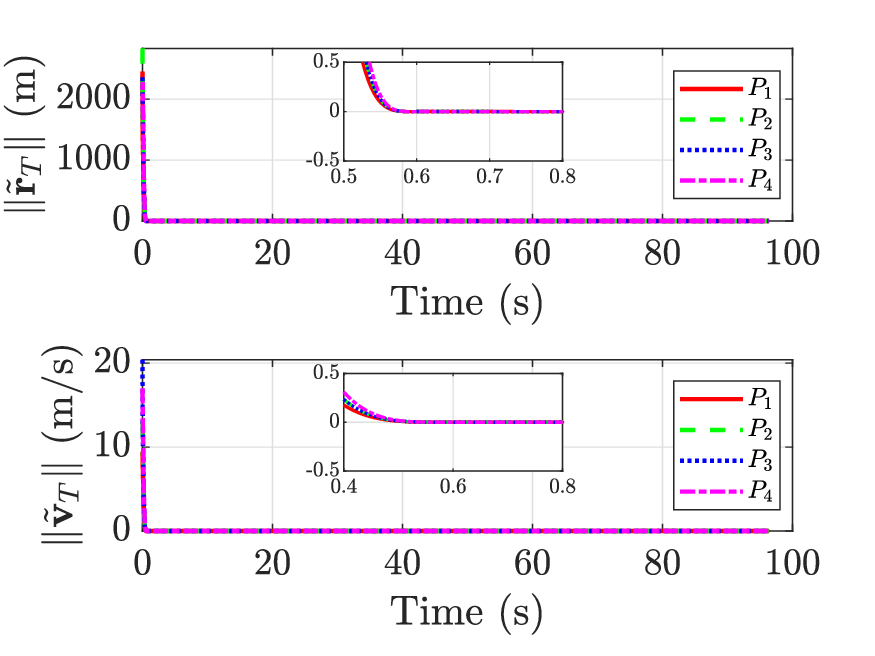}
        \caption{Observer Estimation Error.}
        \label{fig:Observer_errortpn2}
    \end{subfigure}
    
    \begin{subfigure}{0.23\textwidth}
        \centering
        \includegraphics[width=\linewidth]{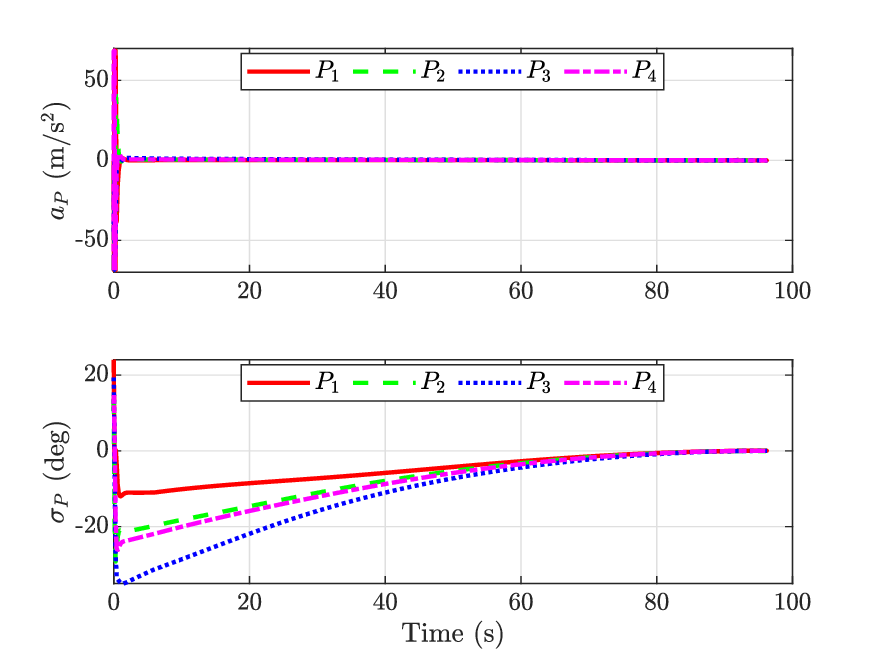}
        \caption{Acceleration \& Lead Angle.}
        \label{fig:accelerationtpn2}
    \end{subfigure}
    \begin{subfigure}{0.23\textwidth}
        \centering
        \includegraphics[width=\linewidth]{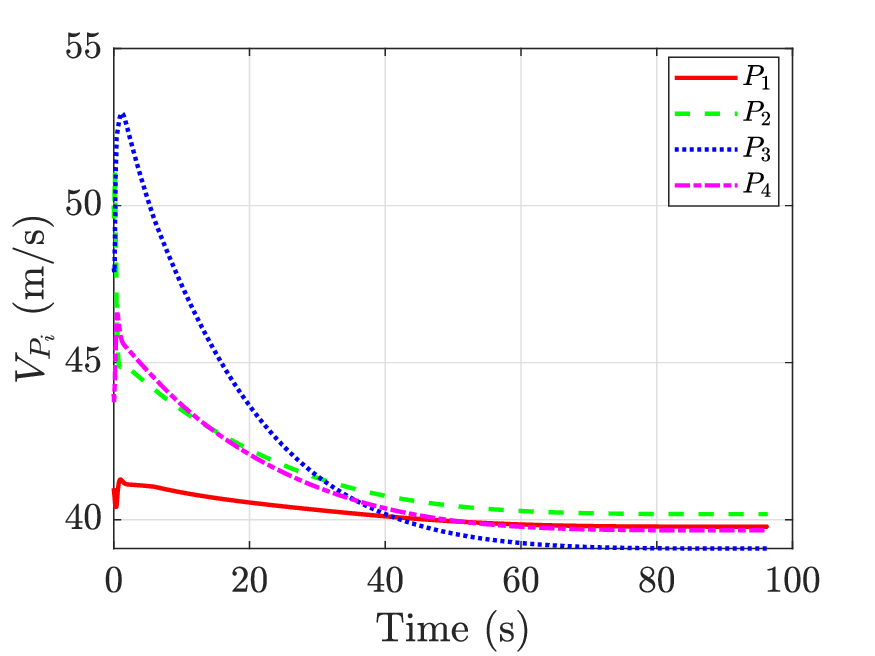}
        \caption{Velocity Evolution.}
        \label{fig:veltpn2}
    \end{subfigure}
    
    \caption{Performance evaluation of the proposed strategy for a stationary target.}
    \label{fig:tpn2}
\end{figure}

To further illustrate the applicability of the proposed framework across a wide range of target motion scenarios, a case involving a stationary target located at  $(3000,2500)$ m is considered. The pursuer speeds are changed to $\begin{bmatrix} 41 & 50 & 48 & 44 \end{bmatrix}$ m, 
with initial heading angles 
$\begin{bmatrix} 25^\circ & 15^\circ & 15^\circ & 20^\circ \end{bmatrix}$ 
and are launched from
$\begin{bmatrix} (75,0) & (0,0) & (150,-40) & (100,-50) \end{bmatrix}$ m. 
The maximum lateral acceleration, observer and controller gains, and observer convergence time are retained from the previous case, while the controller convergence duration is set to $6$ s. The initial target position estimates are taken to be the same as in the previous scenario 
with corresponding velocity estimates of 
$\begin{bmatrix} (5,8) & (10,8) & (20,4) & (8,15) \end{bmatrix}$ m. 
Based on these estimates, the initial time-to-go estimates are 
$\begin{bmatrix} 92.909 & 117.378 & 96.258 & 103.902 \end{bmatrix}$ s, 
while the corresponding true values are 
$\begin{bmatrix} 98.859 & 89.447 & 84.485 & 96.631 \end{bmatrix}$ s.

\Cref{fig:tpn2} depicts the scenario of pursuers achieving cooperative simultaneous interception of a stationary target at $95.9$ s. The corresponding trajectories are shown in \Cref{fig:Trajectorytpn2}. The evolution of time-to-go estimates reaching consensus at $T_a = 6$ s is illustrated in \Cref{fig:Time-to-gotpn2}. The accelerations and lead angles are depicted in \Cref{fig:accelerationtpn2}, while the observer error convergence to zero within $T_o = 0.6$ s is shown in \Cref{fig:Observer_errortpn2}. Similar to the previous case, the lead angles undergo fluctuations, and the acceleration demands remain relatively large before consensus on time-to-go is established. Once consensus is attained at $T_a = 6$ s, both the lead angles and the accelerations are stabilized. In a similar manner, \Cref{fig:veltpn3} illustrates an analogous trend in the velocity evolution, reflecting the same characteristics observed in the previous cases.
\section{Conclusions}
This paper presented a cooperative integrated estimation–guidance framework for simultaneous interception of non-maneuvering targets by unmanned autonomous vehicles with heterogeneous sensing capabilities. The proposed framework was shown to ensure accurate and simultaneous interception under diverse target motions and engagement geometries, which leveraged prescribed-time consensus protocols and true proportional navigation guidance. Simulation results demonstrate the framework’s effectiveness, showing resilience to individual agent failures while maintaining overall coordination. Future work will focus on extending the framework to handle maneuvering targets and implementing time-varying communication topologies to enhance adaptability and operational flexibility.
\bibliographystyle{IEEEtran}
\bibliography{references2,references}

\end{document}